\documentclass[12pt,secnumarabic,showpacs,preprintnumbers,amsmath,amssymb]{revtex4} 
\usepackage{graphicx,epsfig,epsf}% Include figure files 
\usepackage{dcolumn}% Align table columns on decimal point 
\usepackage{bm}% bold math 
\usepackage{slashed} 
\DeclareMathAlphabet{\mathpzc}{OT1}{pzc}{m}{it}

\begin{document}

\title{\bf Semi-classical strings in $(2+1)-$dimensional backgrounds} 
 
\author{Sergio Giardino} 
 \email{giardino@ime.unicamp.br} 
\address{ Instituto de Matem\'{a}tica, Estat\'{i}stica e 
Computa\c{c}\~{a}o Cient\'{i}fica, Universidade Estadual de Campinas\\ 
Rua S\'{e}rgio Buarque de Holanda 651, 13083-859, Campinas, SP, 
Brazil} 
 
\begin{abstract} 
{\bf Abstract} 
This study analyzes the geometrical relationship between a classical 
string and its semi-classical quantum model. From an arbitrary $(2+1)-$dimensional 
geometry, a specific ansatz for a  
classical string is used to generate a semi-classical quantum 
model.  In this framework, examples of quantum oscillations and 
quantum free particles are presented that 
 uniquely determine a classical string and the space-time 
geometry where its motion takes place. 
\end{abstract} 
\noindent  
 
\maketitle 
 
\section{Introduction} 
 
Quantization schemes in string theory are characterized by 
their background dependency. From this standpoint, space-time is something 
more fundamental than the strings, and thus cannot be framed in 
terms of them. This conceptual impediment seems to prevent string 
theory from being  
a quantum model of gravity. However, even if we disregard philosophical 
questions, the technical difficulties in string theory are also great, and a general 
quantization procedure for string theory is as yet unknown.  
 
On the other hand, quantization of string 
theory is possible in specific cases, such as the semi-classical method developed for the 
pulsating string in $AdS_5\times S^5$ 
\cite{Minahan:2002rc,Engquist:2003rn}, a method that has been applied to various 
backgrounds \cite{Giardino:2011jy,Arnaudov:2010by,Arnaudov:2010dk,Beccaria:2010zn,Dimov:2004xi,Smedback:1998yn}. The procedure is  
linked to a particular geometry where classical string motion takes 
place, and in this article a generalization of the method that enables 
a description of a wider class of classical strings in various 
$(2+1)$-dimensional spaces is presented. Strings can be 
understood to move  in an effective space, which is a subspace of the ten 
dimensional space-time where string theory is consistently defined. 
 
The basis for this generalization is the observation that metric 
tensor elements determine the potential of the classical Hamiltonian 
which is used to build a quantum model. From this simple idea, it is 
possible to vary the potential and to establish correspondence between 
the classical model and the quantum model based on the geometry of the 
effective $(2+1)-$dimensional  
space-time where the string moves. The correspondence between the 
quantum model and the classical model does not mean equivalence or duality, as 
in the AdS/CFT correspondence, but an 
association between a classical motion and a quantum model with 
a common space-time geometry. 
It is not clear if there are any other classical strings 
that can generate an identical quantum model, but this question is 
not posed here, because the aim is to demonstrate the existence of  
correspondence only. Potential is the central element relating the 
quantum model to the classical model, and a specific potential requires 
a particular space-time. The results show 
that quantum oscillation and quantum free particles occur at 
different space-time topologies. A space-time that 
could be used to construct a quantum oscillation and a quantum free 
particle for the same string has not been found.

The article is organized as follows: section (\ref{s2}) 
describes dynamics of a string in a $(2+1)-$dimensional and the 
embedding of the string world-sheet in a $3-$dimensional plane 
space. In section (\ref{s3}), string motion and space-time geometry 
are determined from specific potentials of the quantum model, and  
the quantum spectrum is also studied. Section (\ref{s4}) contains the 
the author's conclusions.   
 
\section{the string and space}\label{s2} 
 
 $(2+1)-$dimensional space-time where the string moves is described 
by the line element 
\begin{equation}\label{line} 
ds^2=-dt^2+dx^2+f^2d\varphi^2, 
\end{equation} 
where $x$ is a general coordinate, which can be a distance or an 
angle, $\varphi$ is an angular coordinate and $f$ is 
an arbitrary function of the coordinates. In this space, a classical string performs 
an arbitrary motion. Depending on the string, the equations of motion 
generate the conditions that must be fulfilled by $f$, and the first 
task is to choose a string. 
 
\subsection{classical and quantum dynamics} 
 
The string of interest is described by the ansatz 
\begin{equation}\label{ansatz} 
t=\kappa\tau,\qquad x=x(\tau)\qquad\mbox{and}\qquad\varphi=m\sigma, 
\end{equation} 
where $\kappa$ is a constant and the string is  wrapped $m$ times along 
$\varphi$ and executes some motion along $x$. If $f=f(x)$, the Nambu-Goto action 
for this string, namely 
\begin{equation}\label{nga} 
\mathcal{A}=-m\sqrt{\lambda}\int d\tau\,f\,\sqrt{\kappa^2-\dot{x}^2}, 
\end{equation} 
determines its equation of motion 
\begin{equation}\label{eom} 
\frac{f^\prime}{f}=-\frac{\ddot{x}}{\kappa^2-\dot{x}^2}, 
\end{equation} 
where the prime denotes a derivative with respect to $x$ and the 
dot denotes a derivative with respect to $\tau$. The above equation is 
non-linear and difficult to use, thus it will be substituted by the Virasoro constraint 
\begin{equation}\label{vc} 
\dot{x}^2=\kappa^2-m^2\,f^2. 
\end{equation} 
From (\ref{nga}) and (\ref{vc}), the 
canonical momentum and energy are 
\begin{equation} 
\Pi=\frac{m\sqrt{\lambda}}{\sqrt{\kappa^2-\dot{x}^2}}f\dot{x}\qquad\mbox{and}\qquad \mathpzc{E}=\sqrt{\lambda}\,\kappa 
\end{equation} 
whose canonical Hamiltonian is 
\begin{equation}\label{hamilton} 
\mathcal{H}=\kappa\sqrt{\Pi^2+m^2\lambda\, f^2}. 
\end{equation} 
This classical formalism can be used to semi-classically quantize the 
string using the square of the Hamiltonian (\ref{hamilton}) to express 
the Schr\"{o}dinger equation as 
$\mathcal{H}^2\Psi=\frac{\mathcal{E}^2}{\kappa^2}\Psi$, where 
$\mathcal{E}^2$ is the squared quantum energy. From this analysis, it follows that the coefficient  
$f$ of the line element (\ref{line}) determines the geometry of the 
space as well as  the classical potential and the quantum wave-function.  
 
\subsection{space geometry} 
 
The spatial motion of the string is constrained by the metric (\ref{line}), 
which defines the world-sheet of the string. Consequently, the 
world-sheet of a string moving through the whole of space is identical to 
 space, and thus the embedding of the $2-$dimensional surface 
into a $3-$dimensional plane space enables both  
the geometry of the space and the motion of the string to be visualized. Expressing (\ref{line}) as 
$ds^2=-dt^2+d\mathpzc{s}^2$, the two dimensional 
sub-space generated by $x$ and $\varphi$ must be embedded into a three-dimensional plane 
space with the metric 
\begin{equation} 
 dS^2=g_{ij}\,dx^idx^j=(dx^1)^2+(dx^2)^2+(dx^3)^2. 
\end{equation} 
It is assumed that the coordinates of the plane space have a cylindrical 
symmetry, so that 
\begin{equation}\label{ccs} 
x^1=\rho\sin\phi,\qquad 
x^2=\rho\cos\phi\qquad\mbox{and}\qquad x^3=z, 
\end{equation} 
where $\rho=\rho(x)$, $z=z(x)$ and $\phi=\phi(\varphi)$. The 
embedding is obtained through the identity 
\begin{equation} 
d\mathpzc{s}^2=h_{ab}\,d\xi^ad\xi^b=g_{ij}\frac{\partial{x^i}}{\partial{\xi^a}}\frac{\partial{x^j}}{\partial{\xi^b}}\,d\xi^ad\xi^b,\qquad\mbox{where}\qquad (x,\,\vartheta)=(\xi^1,\,\xi^2). 
\end{equation} 
Using (\ref{ccs}) in the metric tensor $h_{ab}$, we obtained the 
system of differential equations 
\begin{equation}\label{imbed} 
\rho_x^2+z_x^2=1\qquad\mbox{and}\qquad \rho^2\phi_\varphi^2=f^2, 
\end{equation} 
where indices $x$ and $\varphi$ represent derivatives with respect 
to these coordinates. It can immediately be seen that 
$\phi=C\,\varphi$, and as both of the coordinates have the same range, 
 $C=1$ and $\rho=f$. As $f$ is known from the beginning, the embedding must be 
obtained from the integration  
\begin{equation}\label{zeta} 
z=\pm\int\,dx\sqrt{1-f^{\prime\, 2}}. 
\end{equation} 
 
\section{moving strings}\label{s3} 
 
In the preceding section the general formalism which associates a 
moving string in a $(2+1)-$dimensional space to a quantum model has 
been presented. In 
order to relate the geometry and the topology of  
the space to the classical motion of the string, $f$ and $x$ are 
specialized, and then we expect some quantum-geometry 
relation to arise. Maintaining a rotational symmetry for $\varphi$, 
$x$ may be either a radial or an angular coordinate. For a radial 
$x=r>0$, the motion of the string is constrained to an open surface, and 
for an angular coordinate, the surface can be closed. However, this 
division does not exhaust the possibilities, because these two categories 
can be subdivided. Below, the radial coordinate case has 
been divided into the pulsating string case and the falling string 
case. 
 
\subsection{radial coordinate pulsating string} 
In this case, it will be used the ansatz $f=\ell \,r^{\frac{n}{2}}$, where 
$n$  a positive integer and $\ell$ is a dimensional constant 
responsible for $f$ having length dimension. By choosing $y=\left(\frac{m\sqrt{\lambda}\,f}{\kappa}\right)^2$ and 
$\kappa\tau=r\,g(y)$, we obtain the equation of motion from (\ref{vc}) 
\begin{equation}\label{tau} 
g+ n\,y\,g_y=\frac{1}{\sqrt{1-y}}. 
\end{equation} 
Using the hypergeometric function relation 
$F(a,\,b;\,b;\,z)=(1-z)^{-a}$ and contiguous hypergeometric function relations, 
it is possible to ascertain that the general solution for (\ref{tau}) is 
\begin{equation}\label{hyper_pulsating} 
\kappa\tau=r\,F\left(\frac{1}{2},\,\frac{1}{n};\,1+\frac{1}{n};\,\frac{m^2\lambda 
\,\ell}{\kappa^2}r^n\right). 
\end{equation} 
 
For $n=1$, $\kappa\tau\propto\sqrt{1-y}$, which implies that the time 
is limited; hence, this can be discarded as a physical solution. The 
inverse of (\ref{hyper_pulsating}) gives $r=r(\tau)$ for each 
$n$. Although this inverse function is unknown, some considerations 
can be stated by observing the figure 1 below. 
 
For $n=2$, $t=\arcsin(y)$, and thus periodic oscillatory 
behavior for $r$ is warranted. For other values of $n$, the graph shows that 
the maximum value of $y$ is one and that the maximum of $y$ is 
reached at a value of 
$t$, which approaches one the greater $n$. For an even number $n$, 
 $t$ is an odd function where negative values are allowed for the 
argument, indicating a range of $y$ in the interval $[-1,\,1]$ for the 
function. This fact may be observed in figure 2 by formally inverting the first terms of the 
infinite series generated by (\ref{hyper_pulsating}). The same fact 
cannot be stated for odd $n$, where $t$ does not have a definite parity 
for the negative values of the argument. On the other hand, the inversion 
of the first terms confirms the existence of the maximum value of $r$ 
to be one in figure 3. However, as the negative values are not actually allowed in both of 
the cases in this particular problem, the behavior of $r(\tau)$ given by the 
positive $y(t)$ can be described as oscillatory. Of 
course, at $y=0$ the string changes its direction and the 
derivative of $y(t)$ is not defined there. Only when negative values of $y$ are allowed 
for even values of $n$ is the derivative of $y(t)$ well defined 
at these points. The conclusion is that the classical motion 
of $r(\tau)$ is possibly oscillatory for any $n\geq 2$. 
 
Another aspect of the problem to be considered is the geometry of 
the two dimensional surface where the string moves. Defining the 
variable $\,w=f_r^2=\frac{n^2\ell^2}{4}r^{n-2}\,$ and the coordinate 
$z=r\,h(w)$, from (\ref{zeta}) we obtain 
\begin{equation}\label{dablio} 
h+(n-2)w\,h_w=\sqrt{1-w}, 
\end{equation} 
and consequently 
\begin{equation}\label{surface_hyper} 
z=r\,F\left(-\frac{1}{2},\,\frac{1}{n-2};\,1+\frac{1}{n-2};\,\frac{n^2\ell^2}{4}r^{n-2}\right). 
\end{equation} 
(\ref{surface_hyper}) is valid for $n>2$. If 
$n=1$, $w\propto 1/r$, so that $r>1$ and as the string vibrates for 
$r<1$, there is no physical solution for (\ref{dablio}). For $n=2$, 
$z$ is constant, and the whole plane is allowed for  vibration of the 
string. This is an 
expected result coherent with the known solutions 
\cite{Minahan:2002rc,Giardino:2011jy}. The graph of  
(\ref{surface_hyper}) in figure 4 gives an idea of the surface where the string 
is allowed to move 
 
Of course, for each particular $n$, the above graph has an identical reflected line in the 
negative $z$ direction. Besides this, the surface is cylindrically 
symmetric, so that the whole surface is similar to a cone. As $n$ 
increases, the graph approaches a straight line and the 
whole surface approaches a rectangular cone. It is also interesting to note that 
the space is finite. Only the $n=2$ case generates an infinite plane 
surface.  
 
The description of the space and the motion of the string concludes 
the analysis of the classical behavior of the string. The next goal is 
semi-classically study the quantum features of this system. For each 
$n$, a specific quantum model can be obtained, and it is described by the 
Scr\"{o}dinger equation 
\begin{equation}\label{schrod_pulsating} 
-\Psi^{\prime\prime}-\frac{n}{2r}\Psi^{\prime}+m^2\lambda \ell^2\, r^n\Psi=\frac{\mathcal{E}^2}{\kappa^2}\Psi. 
\end{equation} 
Except for the non-physical $n=1$ case, the exactly solvable solution of 
(\ref{schrod_pulsating}) only occurs when $n=2$. For this particular 
situation, the wave-function and the energy spectrum are 
\begin{equation} 
\Psi_N=\mathcal{N}\,e^{-\frac{m\sqrt{\lambda}\,\ell}{2}r^2}L_N(m\sqrt{\lambda}\,\ell\,r^2),\qquad\mbox{and}\qquad \frac{\mathcal{E}^2_N}{\kappa^2m\sqrt{\lambda}\,\ell}=4N+2, 
\end{equation} 
where $\mathcal{N}$ is the normalization constant, $N$ is a positive 
integer and $L_N$ are the Laguerre polynomials. In this case, $\ell$ 
is dimensionless and can be set to one. 
 
The $n>2$ solutions must be studied perturbatively, and the 
non-perturbed case is calculated by excluding the potential term of 
(\ref{schrod_pulsating}). The solutions are 
\begin{equation}\label{free_puls} 
\Psi=\frac{1}{r^{\frac{n-2}{4}}}\Big[A\,J_{\frac{n-2}{4}}\Big(\frac{\mathcal{E}}{\kappa}\,r\Big)+B\,Y_{\frac{n-2}{4}}\Big(\frac{\mathcal{E}}{\kappa}\,r\Big)\Big], 
\end{equation} 
where $A$ and $B$ are integration constants. The space is finite and 
the range of the radial coordinate can be assumed to be 
$[0,\,R]$. Solution (\ref{free_puls}) describes quantum free 
particles, however, the permitted energies can be either continuous 
and quantized. Wave-functions where $\Psi(r=R)\neq 0$ have 
continuous energies and wave-functions where $\Psi(r=R)=0$ have 
a quantized energy spectrum given according to the zeros of the Bessel 
functions,  
\begin{equation}\label{energy} 
\frac{\mathcal{E}_N}{\kappa}=\frac{R^{(N)}_{\mathcal{Z}}}{R},\qquad \mbox{so that}\qquad 
N\in\mathbb{N}\qquad \mbox{and}\qquad \mathcal{Z}=\{J,\,Y\}. 
\end{equation} 
The index $\mathcal{Z}$ is due to the fact that the zeros of the Bessel 
functions $J$ and $Y$ are not common and then there are two independent 
wave-functions and two energy spectra. Both of the 
wave-functions are normalizable, except only the $Y$ wave function 
for $n=6,\,10,\,14,\;etc$.  
 
The perturbative calculations for energy need a wave-function 
given by an orthogonal set, 
and thus only the quantized energy wave-functions can be used. The 
orthogonal set can be obtained from Bessel function $J$, which obeys the condition 
\begin{equation}\label{ortho_puls} 
\intop_0^1dx\,x\,J_\mu(\alpha\,x)\,J_\mu(\beta\,x)=\frac{\beta \,J_\mu(\alpha)\,J_{\mu-1}(\beta)-\alpha\,J_{\mu-1}(\alpha)\,J_\mu(\beta)}{\alpha^2-\beta^2}, 
\end{equation} 
which is zero if $\alpha$ and $\beta$ are different Bessel function 
zeros. From (\ref{ortho_puls}) we also get the normalization 
condition 
\begin{equation} 
\intop_0^1dx\,x\,J^2_\mu\big(R^{(N)}\,x\big)=\frac{J^2_{\mu+1}\big(R^{(N)}_J\big)}{2R^{{(N)}2}_J}. 
\end{equation} 
Thus, the energy in the first order of perturbation for the potential 
$f^2=\ell^2\, r^n$ is  
\begin{eqnarray}\label{correction_puls} 
\frac{\delta\mathcal{E}^2_N}{\kappa^2m^{2}\lambda\,\ell^{2}}&=&\,R^{{(N)}n+1}\intop_0^1dx\,x^{n+1}J^2_{\frac{n-2}{2}}\big(R^{(N)}\,x\big)\nonumber\\ 
&=&\frac{\lambda\,R^{{(N)}2n+2}}{\;n\,\Gamma\left(\frac{n}{2}\right)\,2^{n-2}J^2_{\frac{n-2}{2}}\big(R^{(N)}\big)\;}\, {}_2F_3\Big([n,\frac{n-1}{2}],[n-1,\frac{n}{2},n+1],-R^{{(N)}2}\Big), 
\end{eqnarray} 
where ${}_2F_3$ is a generalized hypergeometric function. As the 
series defined by this object converges for every finite argument,  
 (\ref{correction_puls}) is expected to be a well-behaved value  
that does not diverge for any zero of $J$.  
 
The result rounds off the analysis, which comprises the geometrical 
correspondence between the classical 
dynamics and quantum dynamics of a string. Of course, there is no 
correspondence in the terms of  
gauge/gravity duality, as the classical string does not have a 
quantized spectrum and so the models are not identical in this 
sense. However, the example does show that a classical pulsating string and a 
quantum oscillation are connected through a specific geometry, which 
determines the string motion and the quantum energy spectrum. Another 
example of this correspondence is provided in the next section. 
 
\subsection{free falling string} 
This model is constructed using $f^2=\ell^2\,r^{-n}$, with 
$n\in\mathbb{N}$, and the analysis follows the manner developed for the 
pulsating string, comprising of the classical string motion, the geometry of the 
space and semi-classical quantization. 
 
Choosing $y=\left(\frac{m\,\sqrt{\lambda\ell}f}{\kappa}\right)^2$ and 
$\kappa\tau=r\,g(y)$,  we ascertain from (\ref{vc}) that the 
classical motion obeys  
\begin{equation}\label{tau_2} 
g- n\,y\,g_y=\frac{1}{\sqrt{1-y}}. 
\end{equation} 
From the hypergeometric function relation 
$F(a,\,b;\,b;\,z)=(1-z)^{-a}$ and the contiguous hypergeometric function relations, 
it is possible to ascertain that the general solutions for (\ref{tau_2}) are 
\begin{equation} 
\kappa\tau=r\,F\left(\frac{1}{2},\,-\frac{1}{n};\,1-\frac{1}{n};\,\frac{n^2\,\lambda\,\ell^2}{\kappa^2}r^{-n}\right). 
\end{equation} 
This solution holds for $n>1$, because $n=1$ allows negative values 
for $\tau$, thus comprising an unphysical solution. These solutions 
are different from the former pulsating case, because the 
radial coordinate and the time coordinate continuously increases, as 
can be seen in the figure 5. 
 
The greater the $n$ value, the more the solutions approach the straight 
line  $\kappa\tau=r$. This string goes continuously to infinity, asymptotically 
approaching a constant velocity of a free particle, hence 
it can be described as a string in free fall. 
 
The other aspect of the classical picture, the geometry of the space, 
is obtained from (\ref{zeta}), in a same manner that it was obtained for 
the pulsating string case, and it is described by   
\begin{equation} 
z=\pm r\,F\left(-\frac{1}{2},\,-\frac{1}{n+2};\,1-\frac{1}{n+2};\,\frac{n^2\ell^2}{4}r^{-n-2}\right), 
\end{equation} 
whose positive part can be seen in figure 6. 
 
Of course, each surface has a cylindrical symmetry and it consists of 
two infinite sheets with a hole in the center. The existence of the 
hole is naturally predictable, as the metric is not defined at $r=0$, 
and then the puncture in space is expected. The existence of the two sheets 
where the free fall of the string can occur seems somewhat 
surprising. However, this kind of situation has already  
been observed in a sphere \cite{Minahan:2002rc,Giardino:2011jy}, where 
the string independently pulsates in each hemisphere.  
 
After the classical description, the quantum fluctuations are studied 
through the Schr\"{o}dinger equation 
\begin{equation}\label{schrod_fall} 
-\Psi^{\prime\prime}+\frac{n}{2r}\Psi^{\prime}+\frac{m^2\lambda \ell^2}{r^n}\Psi=\frac{\mathcal{E}^2}{\kappa^2}\Psi. 
\end{equation} 
The $n=1$ has already been observed not to have a classical 
physical solution, and then the analysis comprises $n\geq 2$. There is an exact solution for the $n=2$ case, namely 
\begin{equation} 
\Psi=A\,r\,J_a\Big(\frac{\mathcal{E}}{\kappa}\,r\Big)+\,r\,Y_a\Big(\frac{\mathcal{E}}{\kappa}\,r\Big), 
\end{equation} 
where $a=\sqrt{1+m^2\lambda \ell^2}$ $A$ and $B$ are integration 
constants. For $n>2$, the exact solutions are unknown 
and the free particle solutions are very similar to the aforementioned exact 
solution, given by 
\begin{equation} 
\Psi=A\,r^{\frac{n+2}{4}}\,J_{\frac{n+2}{4}}\Big(\frac{\mathcal{E}}{\kappa}\,r\Big)+\,r\,Y_{\frac{n+2}{4}}\Big(\frac{\mathcal{E}}{\kappa}\,r\Big). 
\end{equation} 
Although the intensity of the wave-function increases with $r$ and 
diverges at infinity, the solutions are indeed free particles. One 
manner of visualizing this is to see that it comes from the fact that 
the non normalizable free particle 
solutions define a Dirac delta function and then obey a localization 
condition \cite{Giardino:2012rb}, 
\begin{equation}\label{dirac} 
\intop_{\infty}^{\infty} \Psi\,\Psi^\star\sqrt{-g}\,dx=\delta(x). 
\end{equation} 
The Dirac delta function in terms of Bessel 
functions in a $(d+1)-$dimensional space, 
\begin{equation} 
\delta^{d+1}\big(\epsilon^2-\eta^2\big)=\intop_0^\infty dr\,r\,J_\mu(\epsilon\,r)\,J_\mu(\eta\,r), 
\end{equation} 
and it fits perfectly with (\ref{dirac}). Thus, even the exact solution 
for $n=2$ is a free particle, and for other values of $n$, the same interpretation holds. 
 
As for the previous case, there are continuous and quantized 
energies. The quantized energies obey the condition that the 
wave-function is zero at the edge of space, and then 
(\ref{energy}) is valid in this case also. The energy of the  
$n>2$ cases are calculated using perturbation theory, so that 
\begin{equation}\label{correction_free} 
\frac{\delta\mathcal{E}^2_N}{\kappa^2m^{2}\lambda\,\ell^{2}}=\,R^{{(N)}n+1}\intop_0^1dx\,x^{n+1}J^2_{\frac{n-2}{2}}\big(R^{(N)}\,x\big) 
\end{equation} 
The exact expression of the above integral, given in terms of 
generalized ${}_2F_3$ functions is complicated and not really 
pertinent to this study. However, as in the former case, the series that 
represents the function is convergent for any value of the argument, 
and this is enough to assure a finite correction to the energy. 
 
\section{conclusion}\label{s4} 
In this article, examples of classical strings were presented that 
 can be semi-classically quantized through a well known 
prescription. The examples demonstrate that the classical string and 
its quantum fluctuations are connected through the space  
where the motion takes place. The geometry and the topology of the 
space determine both the classical string and the quantum 
Hamiltonian.  
 
Although the results extend the range of quantum  
models that can be obtained from a string motion, from the point of 
view of the author of this article, it is somewhat frustrating that the 
potential that goes with the inverse of the distance is not permitted 
in the models presented. The string motion that could model the 
relevant physical phenomena described by this potential, namely 
gravity and electromagnetism remains unknown. On the other hand, the 
results are evidence that the link between quantum theory and general 
relativity through geometry seems not to be merely a myth. 
  
\section*{Acknowledgements} 
Sergio Giardino is thankful for the financial support of Capes and for 
the facilities provided by the IFUSP.   
%%%%%%%%%%%%%%%%%%%% 
% 
% 
%     BIBLIOGRAPHY 
% 
% 

%%%%%%%%%%%%%%%%%%%% 
% 
% 
%     FIGURES 
% 
% 
\pagebreak 
%\vspace{20cm} 
\begin{figure} 
\epsfig{file=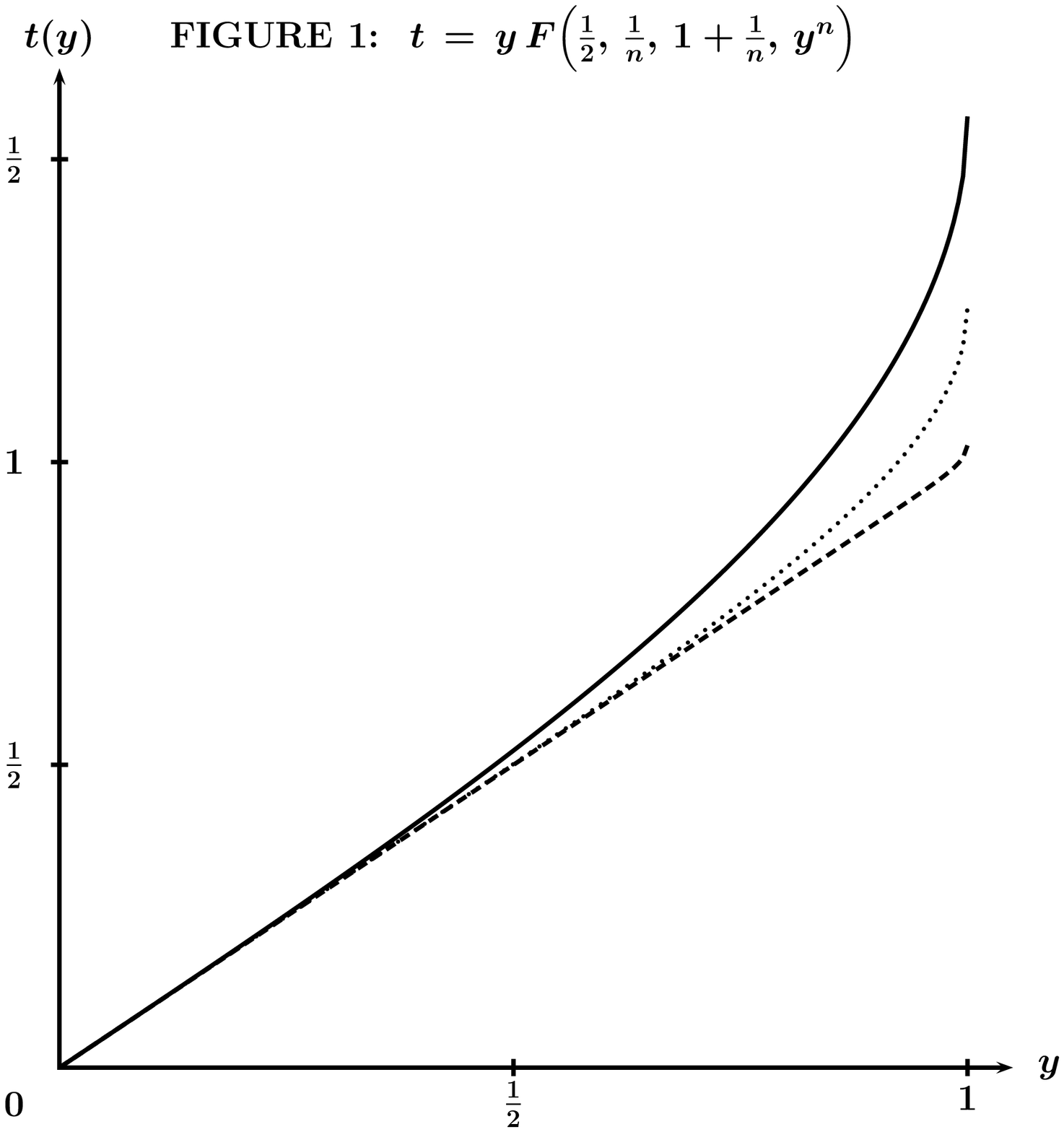,scale=0.8} 
\caption[F1]{\large Time for the pulsating string. 
 
$\qquad\;\;$ solid $n=2\;$ dotted $n=5\;$ dashed $n=50$} 
\end{figure} 
\pagebreak 
\begin{figure} 
\epsfig{file=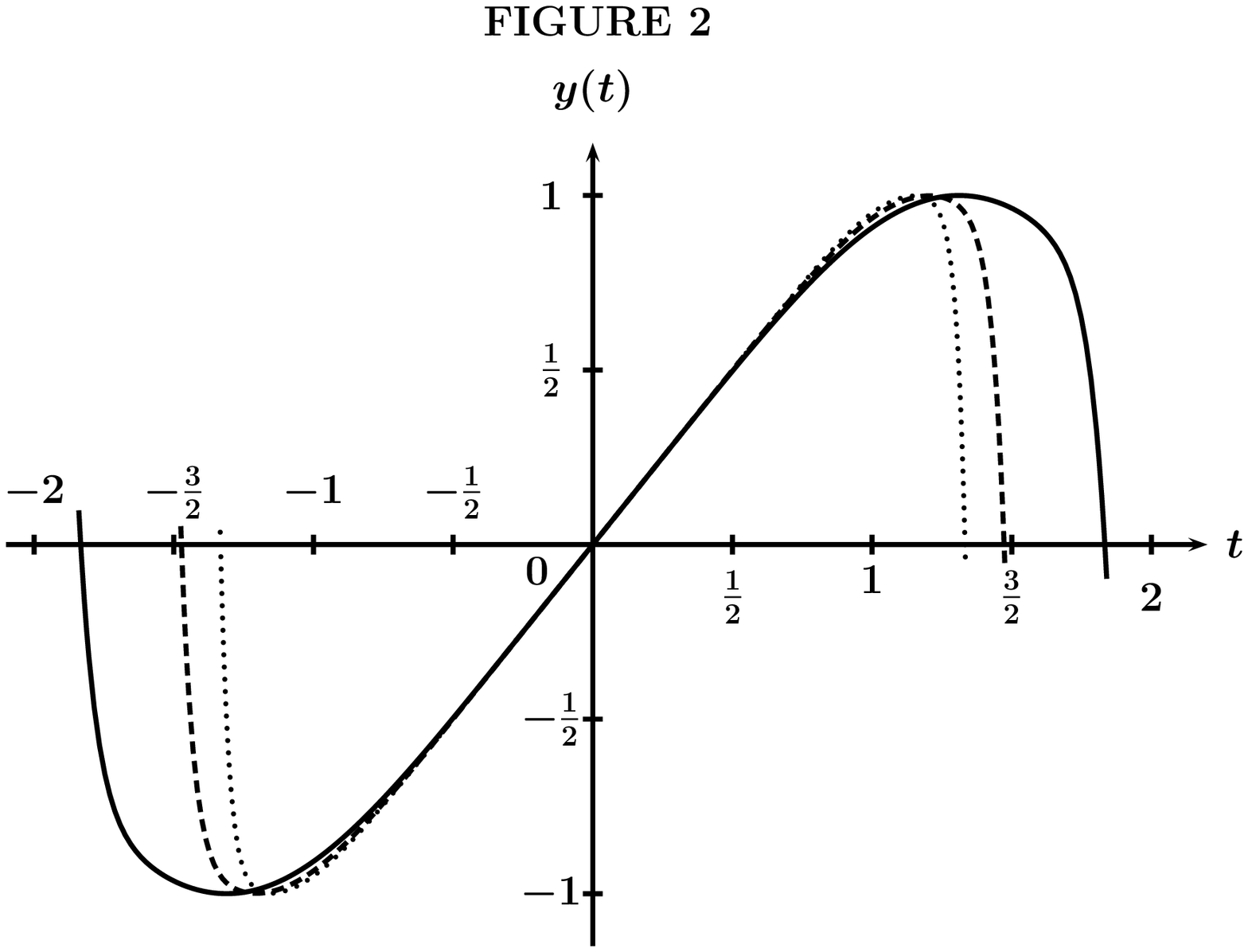,scale=0.8} 
\caption[F2]{\large Formal inversion of $t(y)$. 
 
$\qquad\;\;$ solid $n=2\;$ dashed $n=6\;$ dotted $n=8$} 
\end{figure} 
\pagebreak 
\begin{figure} 
\epsfig{file=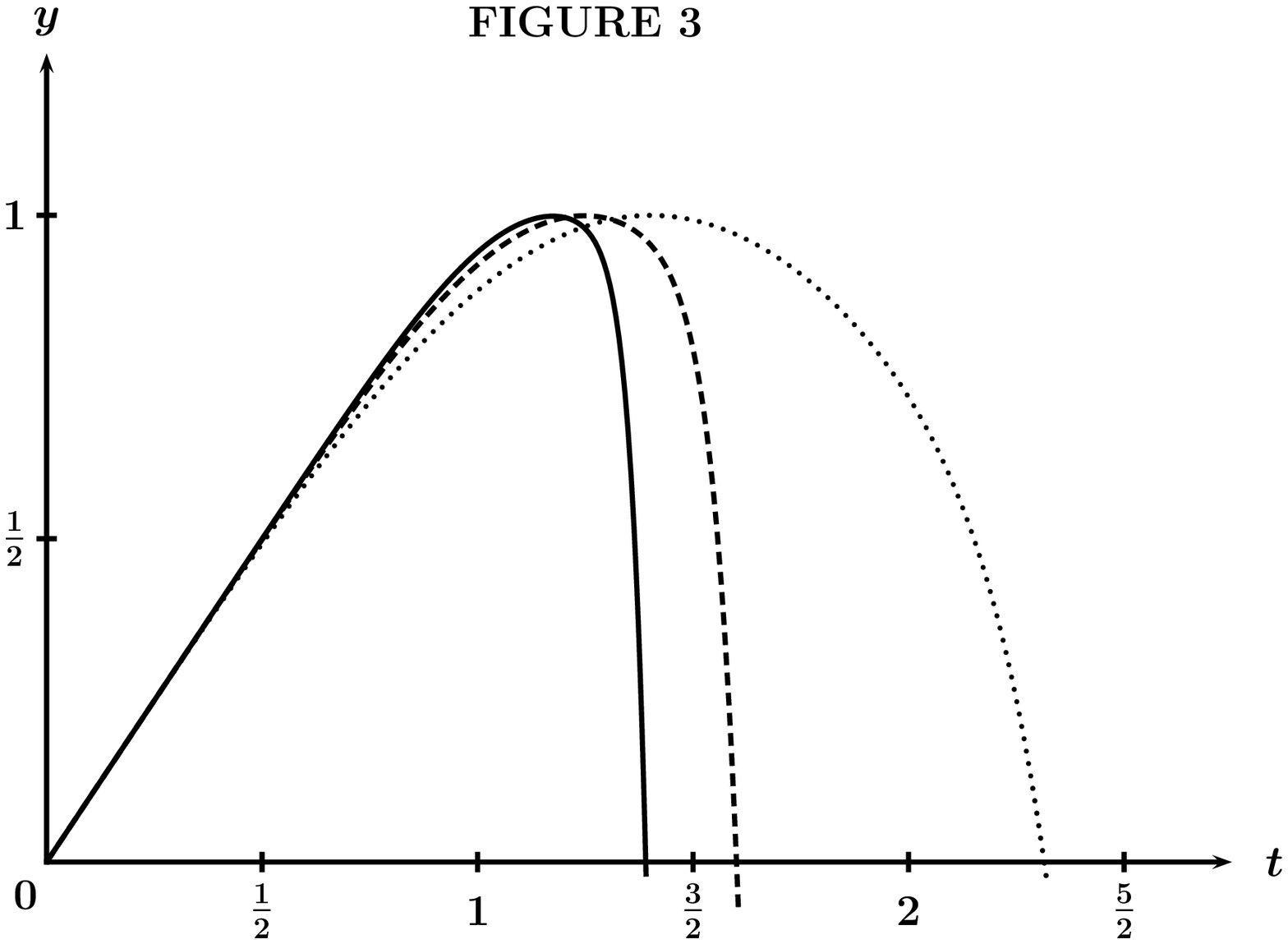,scale=0.8} 
\caption[F3]{\large Formal inversion of $t(y)$. 
 
$\qquad\;\;$ dotted $n=3\;$ dashed $n=5\;$ solid $n=7$} 
\end{figure} 
\pagebreak 
\begin{figure} 
\epsfig{file=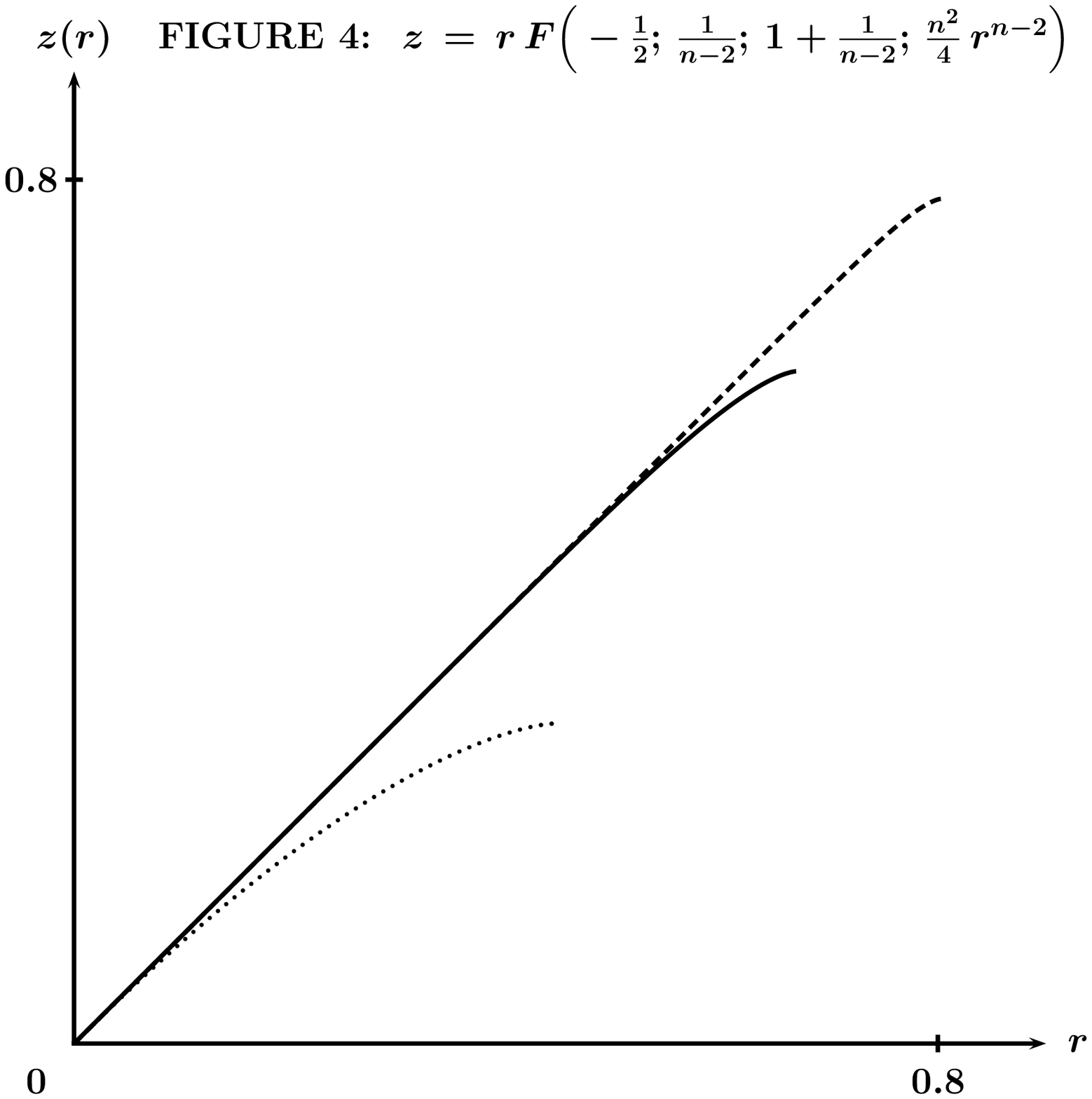,scale=0.8} 
\caption[F4]{\large Embedded space for the pulsating string 
 
$\qquad\;\;$ dotted $n=3\;$ solid $n=10\;$ dashed $n=25$} 
\end{figure} 
\pagebreak 
\begin{figure} 
\epsfig{file=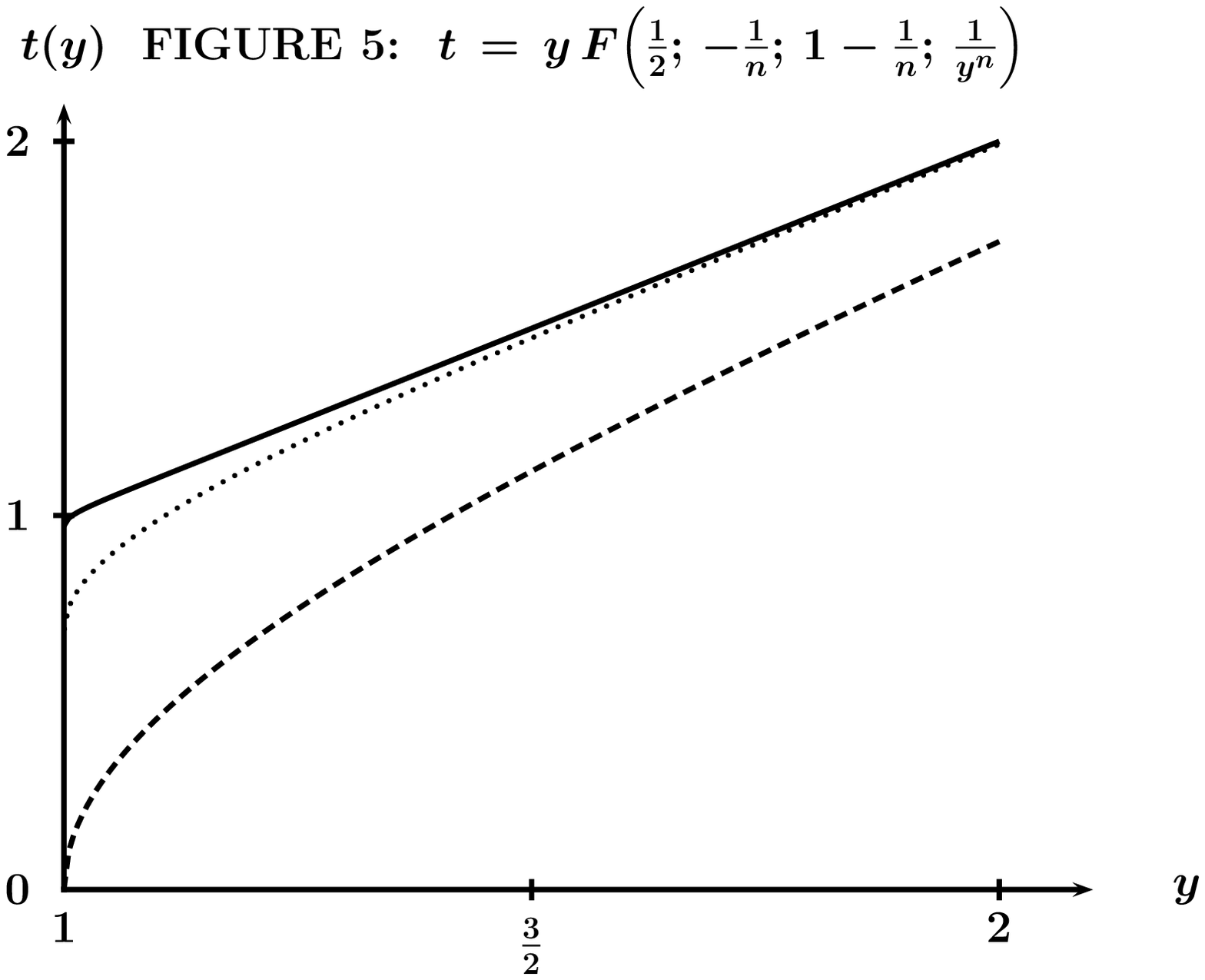,scale=0.8} 
\caption[F5]{\large Time for the free falling string 
 
$\qquad\;\;$ dashed $n=2\;$ dotted $n=5\;$ solid $n=50$} 
\end{figure} 
 
\pagebreak 
\begin{figure} 
\epsfig{file=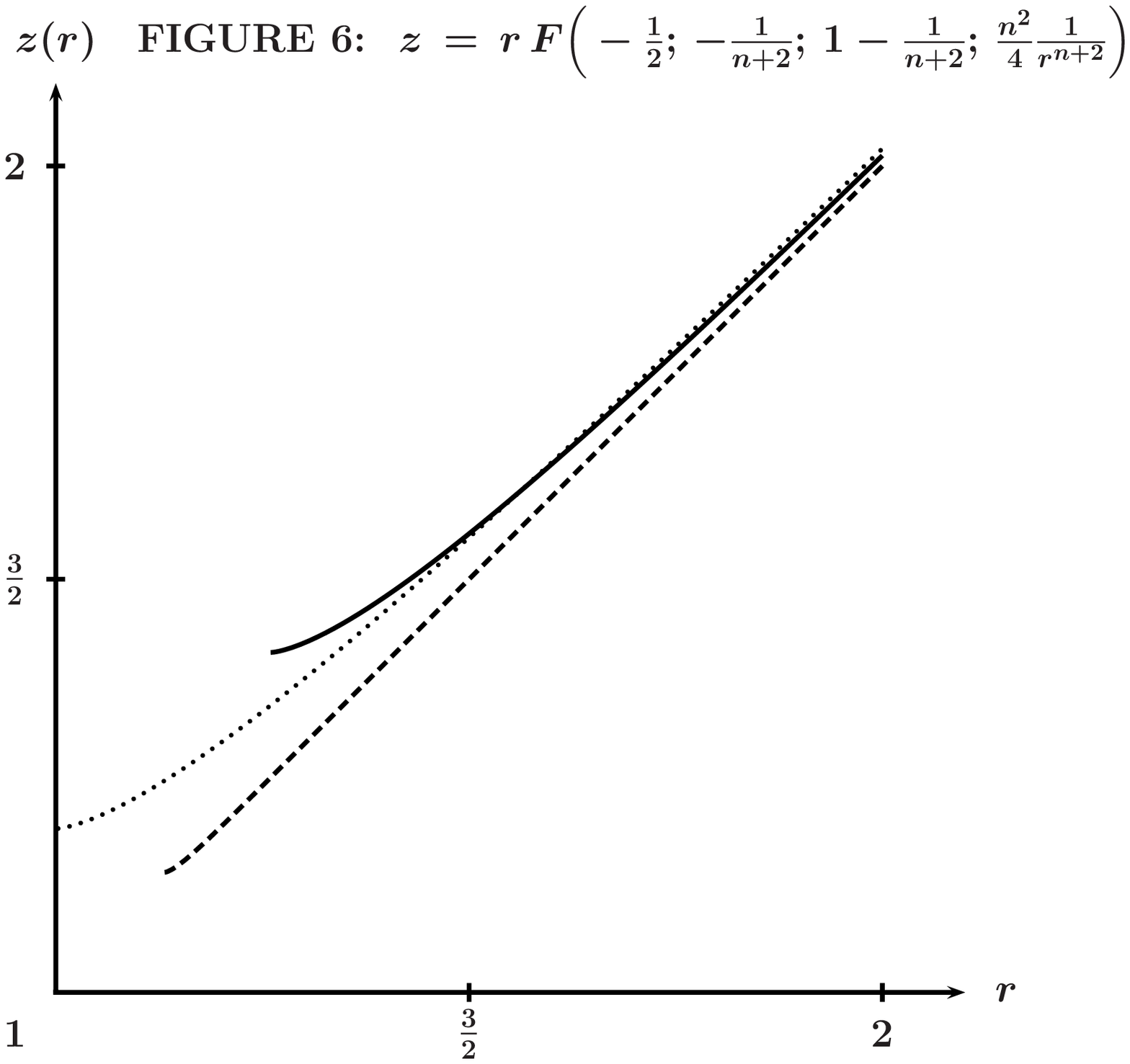,scale=0.8} 
\caption[F6]{\large Embedded space for the free falling string 
 
$\qquad\;\;$ dotted $n=2\;$ solid $n=4\;$ dashed $n=50$} 
\end{figure}

%%%%%%%%%%%%%%%%%% 
% 
% 
%    END 
% 
% 
\end{document}